# Unleashing Modified Deep Learning Models in Efficient COVID-19 Detection

Md Aminul Islam[1], Shabbir Ahmed Shuvo[2], Mohammad Abu Tareq Rony[3], M Raihan[4], and Md Abu Sufian[5]

[1]Oxford Brookes University, UK , atalukder.rana.13@gmail.com,
[2]Offenburg University of Applied Sciences, Germany , shuvo.shabbirahmed@gmail.com,
[3]Noakhali Science and Technology University, Bangladesh , abutareqrony@gmail.com,
[4]Khulna University, Bangladesh, raihanbme@gmail.com
[5]University of Leicester, UK , abusufian.tex.cu@gmail.com

**Abstract.** COVID-19 is a unique and devastating respiratory disease outbreak that has affected global populations as the disease spreads rapidly. Many deep learning breakthroughs may improve COVID-19 prediction and forecasting as a tool for precise and fast detection. In this study, the dataset used contained 8055 CT image samples, 5427 of which were COVID cases and 2628 non-COVID. Again, 9544 X-ray samples included 4044 COVID patients and 5500 non-COVID cases. MobileNet V3 DenseNet201, and GoogleNet Inception V1 show the highest accuracy of 97.872%, and 97.567%, 97.643% respectively. The high accuracy indicates that these models can make many accurate predictions, as well as others, are also high for MobileNetV3 and DenseNet201. An extensive evaluation using accuracy, precision, and recall allows a comprehensive comparison to improve predictive models by combining loss optimization with scalable batch normalization. This research shows that these tactics improve model performance and resilience for advancing COVID-19 prediction and detection and show how Deep Learning can improve disease handling. The methods suggested in this research would strengthen healthcare systems, policymakers, and researchers to make educated decisions to reduce COVID-19 and other contagious diseases.

## 1 Introduction

COVID-19, a novel coronavirus disease that originated in Wuhan, China, in 2019, has exhibited a rapid global spread since the onset of 2020 [1]. It is an infectious ailment characterized by its great transmissibility, resulting from the novel coronavirus 2, which is responsible for inducing severe acute respiratory syndrome (SARS-CoV-2). The contagious nature of the virus is of significant concern, as it possesses frightening attributes that have contributed to the ongoing global pandemic [1]. The presentation of symptoms associated with the condition might vary among individuals, encompassing common manifestations such as fever, headache, dyspnea, dry cough, and anosmia. It is important to note that specific cohorts may exhibit an absence of symptoms in certain instances. It is essential to highlight that those who engage in smoking and drug addiction are particularly susceptible to the impacts of this unique virus. Nevertheless, the inadvertent engagement with individuals in good health within one's vicinity is progressively escalating, contributing to the ailment's transmission rate. The COVID-19 pandemic has been found to harm mortality rates, mental well-being, and human behavioral patterns [2]. Nevertheless, the use of isolation measures stands as a fundamental strategy in combating the transmission of COVID-19. The real-time polymerase chain reaction (RT-PCR) is the prevailing method to identify COVID-19 [3]. Nevertheless, it should be noted that using RT-PCR kits incurs a substantial financial burden, and confirming infection in the patient often requires a considerable time frame of 6-9 hours. The RT-PCR method is associated with a higher likelihood of producing false-negative results due to its reduced sensitivity. In order to address this issue, radiological imaging modalities, such as chest radiography and computed tomography (CT), are employed to identify and diagnose cases of COVID-19. Deep learning-based technologies can expedite the analysis time of chest X-rays using automated means. These methodologies can optimize the weights of neural networks using extensive datasets and adjust the weights of pre-existing networks using limited datasets [3], [4]. Previous studies have demonstrated that biomedical imaging analysis can detect indications of Covid-19. While chest X-rays (CXR) and computed tomography (CT) scans are useful for early detection, radiologists' decision-making can be time-consuming, especially during pandemics and emergency scenarios. Developing and low-income nations with high populations have a disproportionate amount of test samples and radiologists during outbreaks. Automation in diagnostics helps radiologists make faster, more accurate choices. Computer Aided Diagnosis (CAD) technologies diagnose diabetic eye illness, cancer metastases, and Tuberculosis (TB). Machine learning algorithms have helped classify biomedical chest X-rays for pneumonia, TB, and other illnesses. ChexNet shows that deep learning outperforms radiologists. Deep learning algorithms have been used to classify and detect COVID-19 in CXRs, CT scans, and blood samples. Feature extraction and engineering are the main concerns with machine learning. The classifier's efficiency depends on the characteristics used for these tasks, which are often difficult. Deep learning (DL) can automatically obtain ideal features, reducing the requirement for feature extraction and engineering. DL-based medical image classification approaches match expert accuracy. DL-based CAD techniques can outperform radiologists in CT and CXR. Pretrained networks like InceptionV3, VGGNet, InceptionResNetV2, ResNet, and others have successfully detected COVID-19 using CT scan and CXR imaging [5]. There have been prior efforts centered on employing unsupervised learning strategies for this specific endeavor in research. Mittal et al. developed an unsupervised learning method to diagnose COVID-19 using multiple chest imaging modalities. The photographs were divided into two categories, covid and non-covid, using a unique Gravitational Search clustering-based technique. While evaluating the ultrasound dataset, the researchers attained a precision rate of 99.36 percent, whereas their methodology yielded a reduced accuracy of 64.41 percent when applied to the CT dataset. In their study, Rui et al. trained a pulmonary opacity detection model using a limited dataset of COVID-19 CT scan images using unsupervised learning. The researchers reported a 95.5percent detection rate for COVID-19 Multiple object-detection-based methodologies have also been used to detect COVID-19. The authors of the study presented a YOLO-based technique for identifying and distinguishing COVID-19 from other thoracic diseases. The model's detection accuracy was 90.67 percent. In their investigation, Fatima et al. identified objects using a single-shot multi-box detector technique. This technique entails the generation of bounding boxes on chest X-ray images, with each box being assigned a normal or COVID-19 classification label. Researchers have documented a 92 percent classification accuracy rate for COVID-19 identification [6]. The COVID-19 pandemic has achieved global dissemination, affecting all nations across the globe. Furthermore, the prevailing symptoms of this illness remain consistent across affected populations. The impact of the global phenomenon was experienced variably throughout developed and developing nations, with disparities in rates of influence contingent upon their respective economic and social standings. The primary contributions of this study are enumerated as follows:

1. This research created a fusion dataset consisting of combined CT and X-Ray images which is an uncommon approach in deep learning model training for the detection of diseases.
2. This research used thirteen different pre-trained deep learning models modified the final classification layer and replaced it with a series of layers which has immensely improved the classification performance of the models for our dataset.
3. This research introduced sixty freezed layers in the model training process hence using a unique way in the classification model training mix.

## 2  Related Works

The COVID-19 pandemic necessitated accelerated advancements in medical imaging for accurate and swift virus detection. A central methodological pillar in this pursuit has been adopting Deep Learning (DL) techniques, specifically Convolutional Neural Networks (CNNs). Johnson and Martinez (2022) offered a comprehensive examination of CNNs' role in X-ray-based COVID-19 detection. Their findings substantiate the intricate nature of the virus detectable through medical imaging, emphasizing the importance of DL methodologies in the process[7]. Another pivotal approach in this scientific endeavor is the exploration of Transfer Learning (TL). Lee, Kim, and Shin (2023) delved into transfer learning methodologies, specifically in MRI diagnostics for COVID-19. Their study underscored

TL's potential in harnessing pre-trained models, presenting a valuable solution to the challenge of limited labeled training data[8]. Models such as VGG16 and ResNet have been at the forefront of this effort, and their capabilities have been widely recognized for feature extraction in the context of medical imaging. The nuances of these pre-trained models have also been explored. Brown and Patel (2022) probed the efficacy of partially freezing model weights for COVID-19 detection, emphasizing the potential dual benefits: leveraging the pre-trained model's existing knowledge while fine-tuning the specificity of the COVID-19 dataset[9]. Such methodologies mitigate challenges such as underfitting and overfitting, crucial for achieving diagnostic accuracy. However, challenges persist. Roberts and Hayes (2021) raised concerns about potential over-generalization and biases introduced by pre-trained models. They further highlighted the intricacies involved in layer-freezing decisions, pointing toward the gaps in current research and the need for more systematic approaches[10]. Wang and Liu's deep dive in 2023 into the application of CNNs in COVID-19 detection reaffirmed the prominence of CNNs in this domain, accentuating the need for continued innovation and refinement[11]. With rapidly evolving research, it becomes imperative to build upon these findings, addressing existing challenges and driving the frontier of COVID-19 detection through DL and medical imaging. To differentiate between the two illnesses, we have presented the experimental results of the DL model. These results indicate that AlexNet is capable of effectively distinguishing between illnesses based on the images. Furthermore, the model demonstrates superior performance in distinguishing between COVID-19 and lung cancer CT images compared to other scenarios [12]. The authors combined two datasets. Coronal view lung CT scans isolated COVID-19-affected areas. CT scans are analyzed for COVID-19 pneumonia infection using threshold filtering, multi-thresholding, segmentation, and area feature extraction. The study provided a highly accurate CNN model for categorizing individuals as COVID-19 positive or negative. A multi-view fusion model is trained using a deep learning network to assess COVID-19 patients. This model uses CT pictures of the largest lung regions in axial, coronal, and sagittal views. The AUC, accuracy, sensitivity, and specificity are 0.819, 0.760, 0.811, and 0.615. The author used the axial chest computed tomography severity index in this experiment which involves evaluating the photos of 102 participants, 53 males and 49 girls aged 15–79. This study also used visual detection. The Area-Under-Curve value was 89.2 percent, the Sensitivity rate was 83.3 percent, and the Specificity rate was 94 percent using detecting methods. The study examined 81 people. 30 of 81 patients were analyzed using Reverse Transcription Polymerase Chain Reaction (RT-PCR), while 51 were tested using both CTSI and RT-PCR. This study combined RT-PCR laboratory analysis with computed tomography severity index imaging. The sensitivity of COVID-19 detection by computed tomography severity index (CTSI) [13] was 98 percent, but RT-PCR was only 71 percent ($p < 0.001$). A team of Chinese and American radiologists examined 219 patients' data. Again the details of the literature survey are mentioned in Table 1.

**Table 1.** Summary of Previous Study from Literature Review

| Ref. | Model | Data Sources | Type of Dataset | No. Images | Accuracy |
|---|---|---|---|---|---|
| [14] | VGGNet, InceptionNet, ResNet | Public data | Image, Audio, Text | 9540 | 89% |
| [15] | LSA-TransGAN, MedGAN-Net, CNN, VGG16, ResNet50 | COVID-19 (public) | CT images for lungs | 128 synthesized, 86 train set | CNN: 74.55%, VGG16: 75.89%, ResNet50: 78.12% |
| [16] | ResNet50 | PACS of the hospitals | Chest CT scans | 31 images (DICOM format) | 89.9% |
| [17] | Chest X-rays, DenseNet-12 | Hospitals, clinics, radiology departments | Chest X-rays | 112,120 chest X-rays | 94% |
| [18] | CLIP | ILSVRC, CIFAR-10 | MNIST, CIFAR-10, ImageNet | 70,000 training images, 10,000 test images | 73% |
| [19] | - | DenseNet121, ResNet50, VGG16 | Public | CT-COVID19 (44,742 images), CT-NonCOVID19 (1186 images) | 92.3% |
| [20] | Chest X-ray (CXR) | DenseNet121 | In-house dataset | CXR-COVID19 (1000 images), CXR-NonCOVID19 (1000 images) | 95.5% |
| [21] | Chest X-ray (CXR) | ResNet50 | Public | 8000 images | 93.2% |
| [22] | Covid CT, X-ray | AlexNet | Fusion of several datasets (Augmented 3974, Real 1044) | | 94% |
| [12] | X-rays | RestNet101 | - | 8009 images | 82% |

## 3 Methodology

### 3.1 Dataset Preparation

This research has worked with the "Extensive COVID-19 X-Ray and CT Chest Images Dataset" for this research[23]. This dataset consists of non-COVID and COVID cases of both X-ray and CT images. There were 8055 CT image samples in the dataset, of which 5427 were COVID cases and 2628 non-COVID cases. The total number of X-ray samples was 9544, of which 4044 were COVID cases, and 5500 were non-COVID cases. We did not consider X-ray and CT images separately to train our models. Instead, this study created a combined dataset by merging the CT and X-ray images, and after the merge, our combined dataset only contained two classes. The first one is the COVID class consisting of X-ray and CT scan images, and another one is the Non-COVID class, which also contains X-ray and CT scan images. After creating the combined dataset, the train-test split is applied to create a separate train and test set from the dataset. After the split, 80% of the available data for the training set and 20% of the samples for the test set. Then split the train set into train and validation sets where the training set contained 75% of the data samples previously available for the training set. The remaining 25% of the former training set was kept for the validation set. It is also important to mention that this research data augmentation techniques while training the models. This helped the models learn better from the training set. This study resized the images to the size 224*224. Again, random horizontal flips were also applied to the image samples. As pre-trained models are used with frozen layers in our implementation, and the pre-trained models are trained on Imagenet, this study additionally applied Imagenet normalization factors on the images. The Normalization factors we applied in the augmentation with mean values 0.485, 0.456, 0.406, and standard deviations 0.229, 0.224, and 0.225, respectively, across RGB channels.

### 3.2 Model Architectures and Custom Classification Layer Design

In this study, our objective was to evaluate and compare the binary classification performance of many contemporary neural network architectures on our specific dataset. The architectures under consideration were:

- **ResNet variants:** ResNet18, ResNet50, ResNet101
- **DenseNet variants:** DenseNet169, DenseNet201 GoogleNet (Inception V1)
- **MobileNetV3 variants:** MobileNet V3, MobileNet V3 small
- **EfficientNet variants:** EfficientNet B0, EfficientNet B5
- **Other models:** RegNext 2, ShuffleNetV2 x1.0, Deit base patch16 224

The core idea behind our approach was to retain the feature extraction capabilities of these architectures while redesigning the classification component to better fit our specific problem domain Figure 1. To this end, this study took the models mentioned above with pre-trained weights. Then the models were modified by replacing their original classification layers with our custom-designed classifier. This classifier was structured as a sequential neural network comprising the following components:

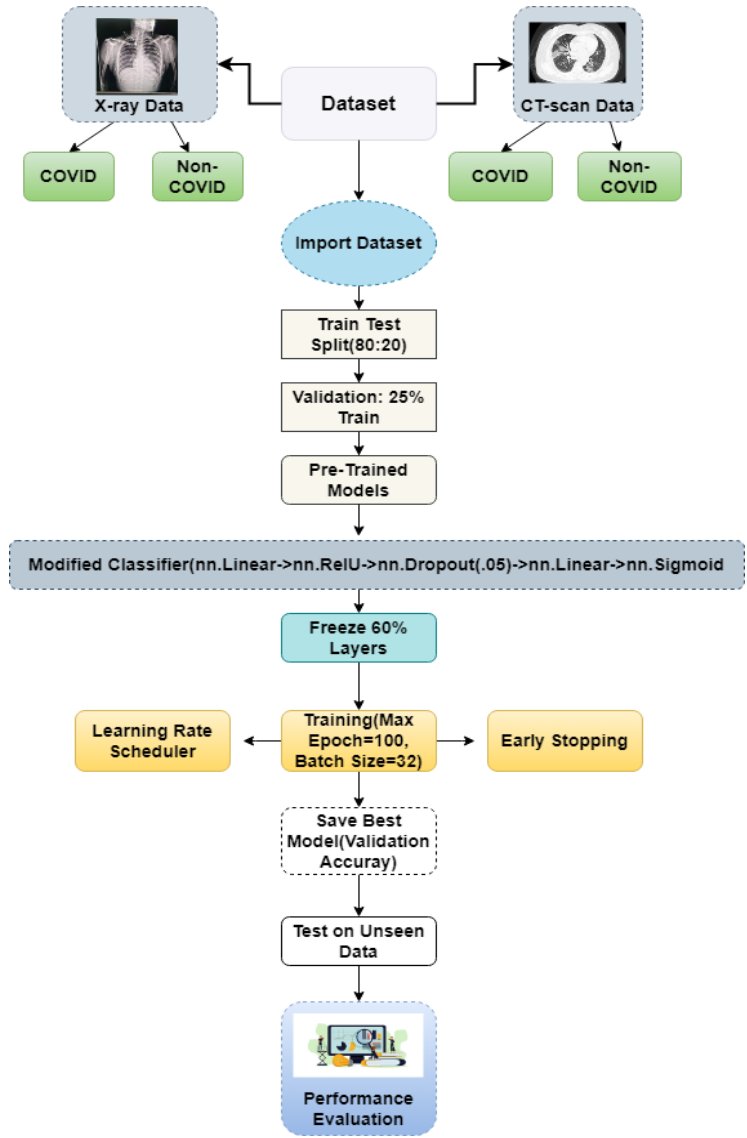

**Fig. 1.** Proposed Methodology

**Linear Layer:** The fully connected (dense) initial layer accepts the feature vectors provided by the pre-trained architectures and transforms them into an intermediate representation. The exact dimensions of this representation were adapted based on the individual architecture's output.

**Rectified Linear Unit (ReLU):** A ReLU activation function was used to add non-linearity after linear transformation. This function ensures that only positive values pass through, allowing the model to capture complex data patterns.

**Dropout:** To mitigate overfitting, a dropout layer was introduced. By stochastically setting a subset of the incoming features to zero during each training iteration, this layer prevents excessive dependence on particular neurons, fostering more generalizable learning.

**Linear Layer:** Another dense layer refines these feature representations, converting them into a scalar output. The dimensionality of this layer was tailored to match each architecture variant.

**Sigmoid Activation:** The output undergoes a sigmoid activation transformation, providing a final output in the [0, 1] range. This makes the result interpretable as the probability of the instance being categorized in a specific class. A custom classification layer like this may provide us with better results. As this author worked with medical images which are much different than normal images, this research could not only rely on pre-trained weights using Imagenet. So we took a transfer learning approach after replacing the classification layer with our custom-designed layer. After preparing the models This study froze 60% of the total layers available for each model in our list. Then in the next stage, This study performed training and testing of the models.

### 3.3  Model Training

After modifying the models and freezing 60% of the total available layers, the models were trained on a training set of the dataset in the next step. The primary validation of the model performance was done on the validation set. We trained the models for a maximum of 100 epochs with batch size 32. We also included best model weight saving, a learning rate scheduler, and early stopping in our training. Given the problem's binary classification nature, the Binary Cross Entropy (BCE) loss was employed as the loss function in the training phase. We utilized the Adam optimizer, a popular adaptive learning rate optimization algorithm to optimize the model's parameters. The Adam optimizer dynamically adjusts the learning rate for each parameter based on the historical gradient moments. We used the default settings for the hyperparameters of Adam, except for the learning rate. Furthermore, the optimization was applied only to those model parameters that required gradients, ensuring efficient and focused parameter updates. model training algorithm used in this study is mentioned in Algorithm 1.

### 3.4  Experimental Setup

We implemented the completed code using Python version 3.9.17. We used the Conda environment and locally run Jupyter instances to build, train, and test our models. We used Pytorch version 2.1.0.dev20230803 to build the models with Cuda version 12.1. Additionally, we used Scikit-learn and Numpy for data loading and processing and Matplotlib and Seaborn for visualization purposes. We trained and evaluated the models on the Ubuntu 22.04.3 LTS operating system on an Intel-based PC with NVIDIA GPU.

## 4  Model Evaluation

As shown in the mentioned training algorithm above in algorithm 1 below, we were saving the model weights of the models with the best performance on the validation set. Once the training was finished, we loaded the best model weights saved while training for evaluation. We evaluated the performance of the loaded model on a test set that consists of data that the trained model has not seen before. For performance comparison with other models, we have considered the following parameters[24]: We have obtained these measurements for each of the considered models for this work and at the end we have compared their values to make important discoveries about the performance of the models on our specific dataset and specific settings. Different evaluation metrics are calculated from the confusion matrix which is a useful tool for visualizing the performance of a model [25] and gaining insights into the occurrences of true positives, false positives, true negatives, and false negatives are provided below :

Accuracy is computed as the ratio of correctly classified instances and the total number of predicted values. This formula is given as follows:

$$\text{Accuracy} = \frac{\text{True Value} + \text{TP} + \text{TN}}{\text{TP} + \text{FP} + \text{TN} + \text{FN}} \quad (1)$$

Precision is a ratio of correctly classified positive instances with a total number of classified positive samples. The formula of this metric is provided as follows;

$$\text{Precision} = \frac{TP}{TP + FP} \quad (2)$$

Recall is a ratio of correctly classified positive instances with the total number of actual positive samples. The formula of this metric is shown as follows:

$$\text{Recall} = \frac{TP}{TP + FN} \quad (3)$$

The harmonic mean of recall and accuracy is the F1-score as follows;

$$\text{F1-score} = \frac{2 \times \text{Recall} \times \text{Precision}}{\text{Recall} + \text{Precision}} \quad (4)$$

Again Receiver Operating Characteristic (ROC) curve is utilized to visually represent the performance of a model across various probability thresholds. Finally, Area Under the Curve (AUC) is calculated to assess the model's performance. The precision-recall curve is generated to analyze the trade-off between precision and recall across different probability thresholds.

---

**Algorithm 1** Model Training
---
**Input**:
Training data *train_loader*
Validation data *val_loader*
Epochs *EPOCHS*
Early stop threshold *PATIENCE*

**Output** Best model parameters

**procedure** MODEL TRAINING
    best_acc = 0.0
    no_improvement = 0
    early_stop = False

    **while** epoch in 1 to *EPOCHS* **do**
      **if** early_stop **then**
        **break**
      **end if**
      **Train**:
      **Mode: Training**
      **for** batch in *train_loader* **do**
        Compute loss, backpropagate, update weights
      **end for**
      **Validate**:
      **Mode: Evaluation**
      **for** batch in *val_loader* **do**
        Compute loss, accuracy
      **end for**
      **Checkpoint**:
      **if** accuracy greater then best_acc **then**
        Save model, reset no_improvement
      **else**
        no_improvement++
      **end if**
      **Early Stopping**:
      **if** no_improvement $\geq$ *PATIENCE* **then**
        early_stop = True
      **end if**
    **end while**
    **End**:
    **Return** best model
**end procedure**

## 5   Result and Discussion

As mentioned before, we trained our models on the training set, validated them on the validation set, and then tested on the test set data from our prepared dataset. All results are results obtained from separate testing sets which the trained model did not see until the test was performed. Based on the results in Table 2 in the above table, we can observe the following: The models with the highest accuracy are MobileNet V3 (97.872%), DenseNet201 (97.567%), and GoogleNet Inception V1 (97.643%). High accuracy indicates that these models make a high proportion of correct predictions. Again, the models with the highest precision are DenseNet201 (98.051%) and RegNext 2 (97.739%). High precision indicates that these models have a low rate of false positives, which is crucial in medical applications where false positives can lead to unnecessary stress and additional testing. Moreover, AThe models with the highest recall are MobileNet V3 (98.704%), ShuffleNetV2 x1.0 (98.704%), and GoogleNet Inception V1 (98.639%). High recall indicates that these models are good at identifying positive cases, essential in a pandemic where early detection is critical. Again, The models with the highest F1 scores are MobileNet V3 (98.195%), RegNext 2 (97.897%), and DenseNet201 (97.923%). A high F1 score indicates a balance between precision and recall, which is desirable in applications where false positives and negatives have significant consequences. The models with the highest ROC-AUC are DenseNet201 (99.702%), RegNext 2 (99.683%), and MobileNet V3 (99.630%). A high ROC-AUC indicates that these models are excellent at distinguishing between positive and negative cases. Considering the performance based on all of the above performance metrics, the Modified DenseNet201 model emerges as the most suitable choice for COVID-19 detection for our specific dataset and use case. The model has the following scores:

**Table 2.** Comparison of All Models

| Modified Model Name | Accuracy (%) | Precision (%) | Recall (%) | F1-Score (%) | ROC-AUC (%) |
|---|---|---|---|---|---|
| ResNet18 | 96.617 | 97.331 | 96.889 | 97.109 | 99.444 |
| ResNet50 | 97.073 | 97.784 | 97.213 | 97.498 | 99.483 |
| ResNet101 | 96.693 | 97.211 | 97.148 | 97.179 | 99.468 |
| DenseNet169 | 97.301 | 97.668 | 97.732 | 97.700 | 99.577 |
| DenseNet201 | 97.567 | 98.051 | 97.797 | 97.923 | 99.702 |
| GoogleNet Inception V1 | 97.643 | 97.377 | 98.639 | 98.004 | 99.624 |
| MobileNet V3 | 97.872 | 97.691 | 98.704 | 98.195 | 99.630 |
| MobileNet V3 small | 96.959 | 97.224 | 97.602 | 97.413 | 99.594 |
| EfficientNet B0 | 96.047 | 96.449 | 96.824 | 96.636 | 99.266 |
| EfficientNet B5 | 95.781 | 95.663 | 97.213 | 96.432 | 99.198 |
| RegNext 2 | 97.529 | 97.739 | 98.056 | 97.897 | 99.683 |
| ShuffleNetV2 x1.0 | 97.263 | 96.698 | 98.704 | 97.691 | 99.619 |
| Deit base_patch16_224 | 82.630 | 88.786 | 80.557 | 84.472 | 93.111 |
| Vit base_patch16_224 | 95.325 | 95.396 | 96.695 | 96.041 | 98.979 |

Considering all these factors, the modified DenseNet201 model emerges as the most suitable choice for COVID-19 detection for our specific dataset and use case. The model has Accuracy: 97.567, Precision: 98.051%, Recall: 97.797%, F1-score: 97.923% and ROC-AUC: 99.702%. It has the highest precision and ROC-AUC among all the models and high values in the other metrics. This indicates that the model is good at making correct predictions and distinguishing between positive and negative cases, which is crucial in a medical context. In the following figures Figure 2 to 5, we can see training vs validation accuracy, training vs validation loss, Confusion metrics, and ROC curve plots for the modified DenseNet201 model, and finally Histogram of predicted probabilities,

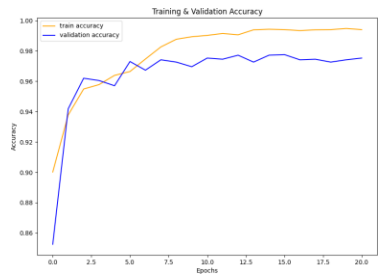
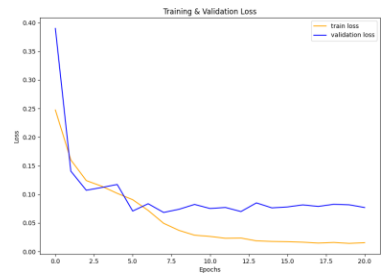

**Fig. 2.** DenseNet201 Accuracy          **Fig. 3.** DenseNet201 Loss

## 6    Conclusion and Future Work

In the context of COVID-19 detection, transfer learning with partially frozen weights appears to be a promising approach. Several deep learning architectures, when fine-tuned with this method, show high efficacy in distinguishing between positive and negative cases. However, the effectiveness varies across architectures, underscoring the importance of robust comparisons to identify the most suitable model for the task. The present study uses deep ensemble models to detect COVID-19 cases, overcoming the cost and time constraints of RT-PCR testing by the Modified DenseNet201 model architectures, outperforming other designs with accuracy 97.567%, precision 98.051%, recall 97.797%, F1-score 97.923%, and ROC-AUC 99.702%. The fluid and evolving symptoms of COVID-19 make it challenging to use exact diagnostic criteria and a vast range of symptoms, which may resemble flu-like symptoms, requires flexibility. The suggested approach surpasses previous studies and achieves high accuracy, unquestionably significant. This research also makes a side-by-side comparison and selects the most reliable classifier for the model. In the future, advanced deep-learning approaches can be used to make our model more robust and reliable. The design ecosystem with advanced classification algorithms can predict many things. The work can be extended to solve real-world medical data and improve analysis automation, including other deep learning algorithms.

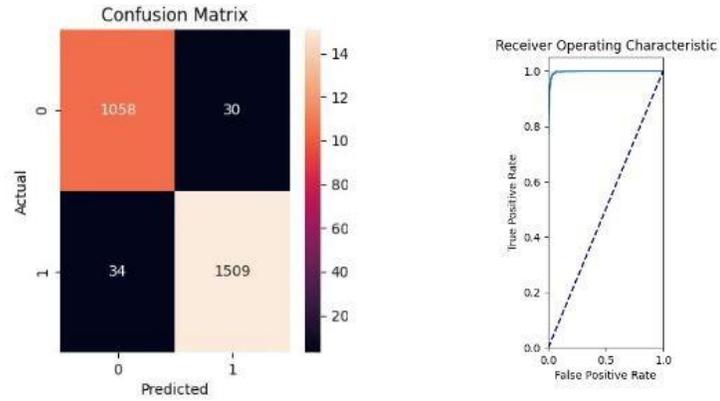

**Fig. 4.** Confusion Matrix and ROC of DenseNet201

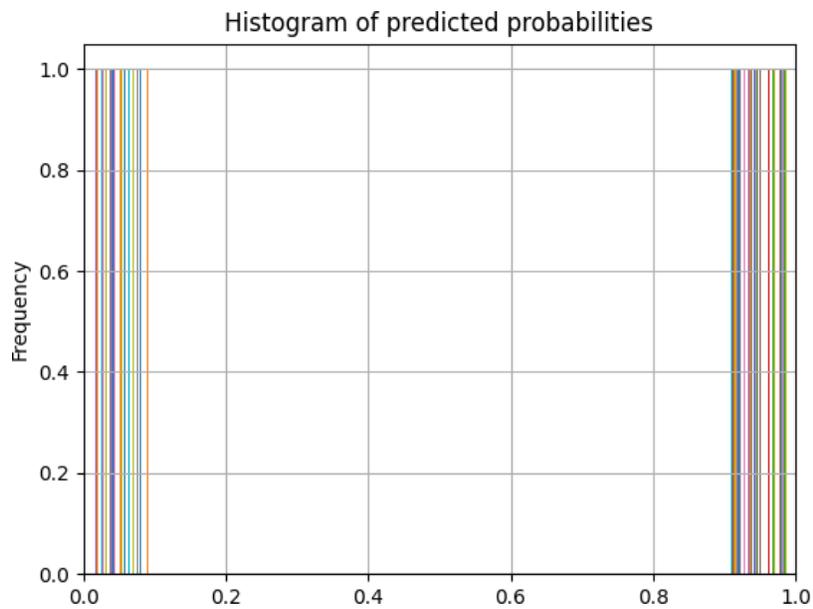

**Fig. 5.** DenseNet201 Histogram

14      Aminul. et al.

Unleashing Modified Deep Learning Models in Efficient COVID-19 Detection     1514. Dipanjan Sarkar, Raghav Bali, and Tamoghna Ghosh. *Hands-On Transfer Learning with Python: Implement advanced deep learning and neural network models using TensorFlow and Keras*. Packt Publishing Ltd, 2018.
15. Shanling Nie, Yunjie Cai, Yue Guo, Zeqi Zheng, and Hai Yang. Medgan-net: a computer-aided diagnosis approach based on deep learning for covid-19 with ct images (preprint). 2022.
16. Zheng Ye, Yun Zhang, Yi Wang, Zixiang Huang, and Bin Song. Chest ct manifestations of new coronavirus disease 2019 (covid-19): a pictorial review. *European radiology*, 30:4381–4389, 2020.
17. Angham G Hadi, Mohammed Kadhom, Nany Hairunisa, Emad Yousif, and Salam A Mohammed. A review on covid-19: origin, spread, symptoms, treatment, and prevention. *Biointerface Research in Applied Chemistry*, 10(6):7234–7242, 2020.
18. Nandhini Subramanian, Omar Elharrouss, Somaya Al-Maadeed, and Muhammed Chowdhury. A review of deep learning-based detection methods for covid-19. *Computers in Biology and Medicine*, 143:105233, 2022.
19. M Rubaiyat Hossain Mondal, Subrato Bharati, and Prajoy Podder. Diagnosis of covid-19 using machine learning and deep learning: a review. *Current Medical Imaging*, 17(12):1403–1418, 2021.
20. Jia Liu, Jing Qi, Wei Chen, Yi Wu, and Yongjian Nian. Deep learning for detecting covid-19 using medical images, 2022.
21. Nandhini Subramanian, Omar Elharrouss, Somaya Al-Maadeed, and Muhammed Chowdhury. A review of deep learning-based detection methods for covid-19. *Computers in Biology and Medicine*, 143:105233, 2022.
22. Falana William, Ali Serener, and Sertan Serte. Effect of multimodal imaging on covid-19 and lung cancer classification via deep learning. In *2021 5th International Symposium on Multidisciplinary Studies and Innovative Technologies (ISMSIT)*, pages 120–124. IEEE, 2021.
23. Walid El-Shafai and Fathi Abd El-Samie. Extensive covid-19 x-ray and ct chest images dataset. *Mendeley data*, 3(10), 2020.
24. Md Aminul Islam, Abu Sufian, and Shabbir Ahmed Shuvo. Data analytics on key indicators for the city's urban services and dashboards for leadership and decision-making. *arXiv preprint arXiv:2212.03081*, 2022.
25. Md Simul Hasan Talukder, Mohammad Raziuddin Chowdhury, Md Sakib Ullah Sourav, Abdullah Al Rakin, Shabbir Ahmed Shuvo, Rejwan Bin Sulaiman, Musarrat Saberin Nipun, Muntarin Islam, Mst Rumpa Islam, Md Aminul Islam, et al. Jutepestdetect: An intelligent approach for jute pest identification using fine-tuned transfer learning. *Smart Agricultural Technology*, 5:100279, 2023.